\documentclass[conference]{IEEEtran}
\usepackage{cite}

\ifCLASSINFOpdf
   \usepackage[pdftex]{graphicx}
   \DeclareGraphicsExtensions{.pdf,.jpeg,.png}
\else
\fi
\begin{document}
%
\title{Kanban + X: Leveraging Kanban for Focused Improvements}

\author{\IEEEauthorblockN{Adam Hey}
\IEEEauthorblockA{St. John's University\\
Collegeville, MN 56321 \\
Email: ajhey@csbsju.com}
\and
\IEEEauthorblockN{Michael A.~Heroux}
\IEEEauthorblockA{St. John's University\\
Collegeville, MN 56321\\
Email: mheroux@csbsju.edu \\
Sandia National Laboratories \\
Email: maherou@sandia.gov}
}


%


\maketitle

\begin{abstract}
Agile Development is used for many problems, often with different priorities and challenges. However, generalized engineering methodologies often overlook the particularities of a project. To solve this problem, we have looked at ways engineers have modified development methodologies for a particular focus, and created a generalized framework for leveraging Kanban towards focused improvements. The result is a parallel iterative board that tracks and visualizes progress towards a focus, which we have applied to security, sustainability, and high performance as examples. Through use of this system, software projects can be more focused and directed towards their goals.
\end{abstract}


%
\IEEEpeerreviewmaketitle

\tableofcontents

\listoffigures

\section{Introduction}
Agile Methodologies are useful for supporting productive development in Software Engineering projects. Some projects, however, have specific focuses that are of high priority for the desired software. In their basic forms, Agile Methodologies may neglect focuses. In these cases, it is useful for the project's success that a focus be leveraged in the development process itself. By looking at a successful example of security-focused Scrum, this paper attempts to apply the same strategy to Kanban and generalize it for any possible focus.

\section{Scrum + Security: Scrum Leveraged for Security}

\begin{figure}
	\centering
	\includegraphics[width=\columnwidth]{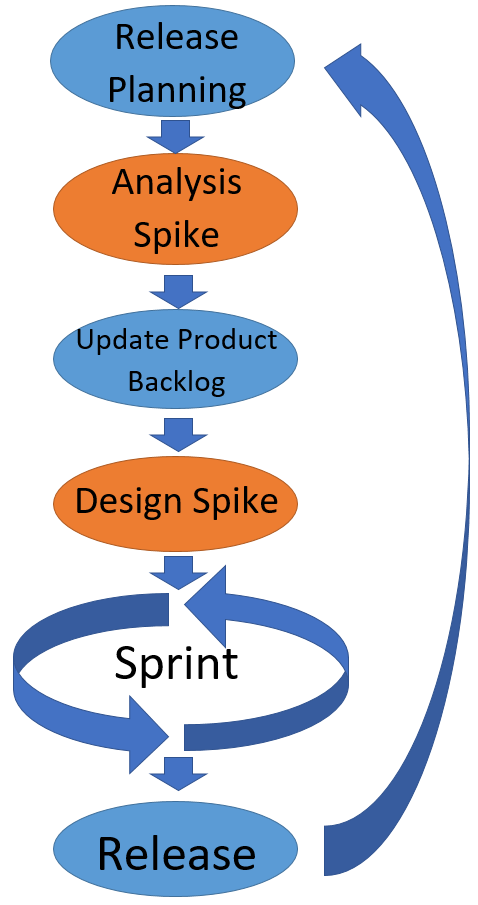}
	\caption[S-Scrum: A Model for Safe Agile Development]{\label{fig:s-scrum}} The S-Scrum work flow. By adding steps for security analysis and security design before the start of each sprint, S-Scrum makes security concerns unavoidable and improves the security of Scrum. The steps are called "Spikes" in accordance with Scrum terminology. After the analysis step, the backlog is modified in accordance with identified security issues \cite{RN68}.
\end{figure}

One successful example of an Agile methodology being modified for a specific focus is the leveraging of Scrum towards Security. Security is a major concern for certain software projects, such as web services \cite{RN68}. However, Scrum, the most popular Agile methodology in the industry, is often criticized for neglecting security analysis and design \cite{RN68}.

In response, researchers at UPM, one of Malaysia's leading research universities, created a modified Scrum that incorporated secure analysis and design into the Agile process \cite{RN68}. They called this secure version of Scrum "S-Scrum"; its goal: to improve security of web service projects using Scrum \cite{RN68}. S-Scrum adds a security analysis and security design step to the Scrum work-flow (see: Figure \ref{fig:s-scrum}). These steps are called "Spikes" in accordance with Scrum terminology. Prior to the start of each sprint, these steps are completed, and the backlog is modified to reflect needed security changes. In this way, the Scrum process is leveraged towards its focus, and security is improved.

This idea of a Security-leveraged Scrum was further refined and evaluated by researchers at the Federal Institute of Education, Science and Technology of Sao Paulo \cite{RN14} and the Munich IT Security Research Group \cite{RN52}. 

\begin{figure*}
	\centering
	\includegraphics[width=35pc]{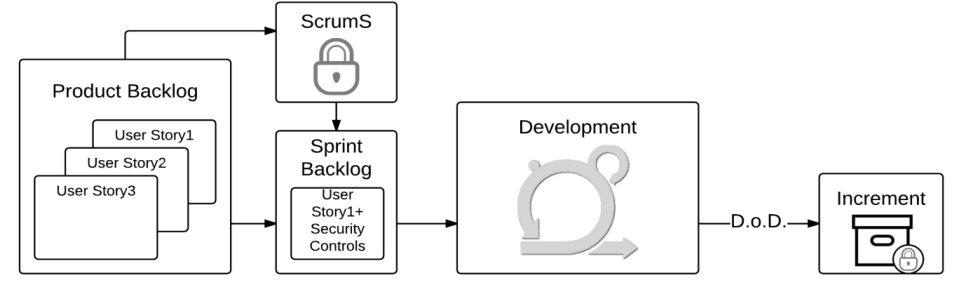}
	\caption[ScrumS: Incorporating Security into Sprints]{\label{fig:scrums} The ScrumS work flow. ScrumS improves S-Scrum by incorporating security analysis and design into Scrum's Sprint cycles. Security-related tasks are extracted from user stories and completed in an iterative process running parallel to primary development. Results are passed on to the sprint backlog for appropriate security modifications. \cite{RN14}}
\end{figure*}

The Sao Paulo research team's revision, "ScrumS", placed the security analysis and design \textit{inside} of the sprint cycle. As tasks are moved from the Product Backlog to the Sprint Backlog, security-related tasks are extracted from ordinary development stories \cite{RN14}. These "Security User Stories" are completed in a parallel iterative loop within the same sprint (see Figure: \ref{fig:scrums}). As Security User Stories are completed, prescribed changes are fed back into the Sprint Backlog, where they are implemented with the rest of the sprint. By integrating security analysis and design into sprints, ScrumS does not slow down the time between iterations. Thus, it better leverages security within the context of Scrum's iterative process. The researchers performed a case study using ScrumS on a security-focused surveying project, and found the resulting software was more secure than previous Scrum projects \cite{RN14}. 

\begin{figure}
	\centering
	\includegraphics[width=\columnwidth]{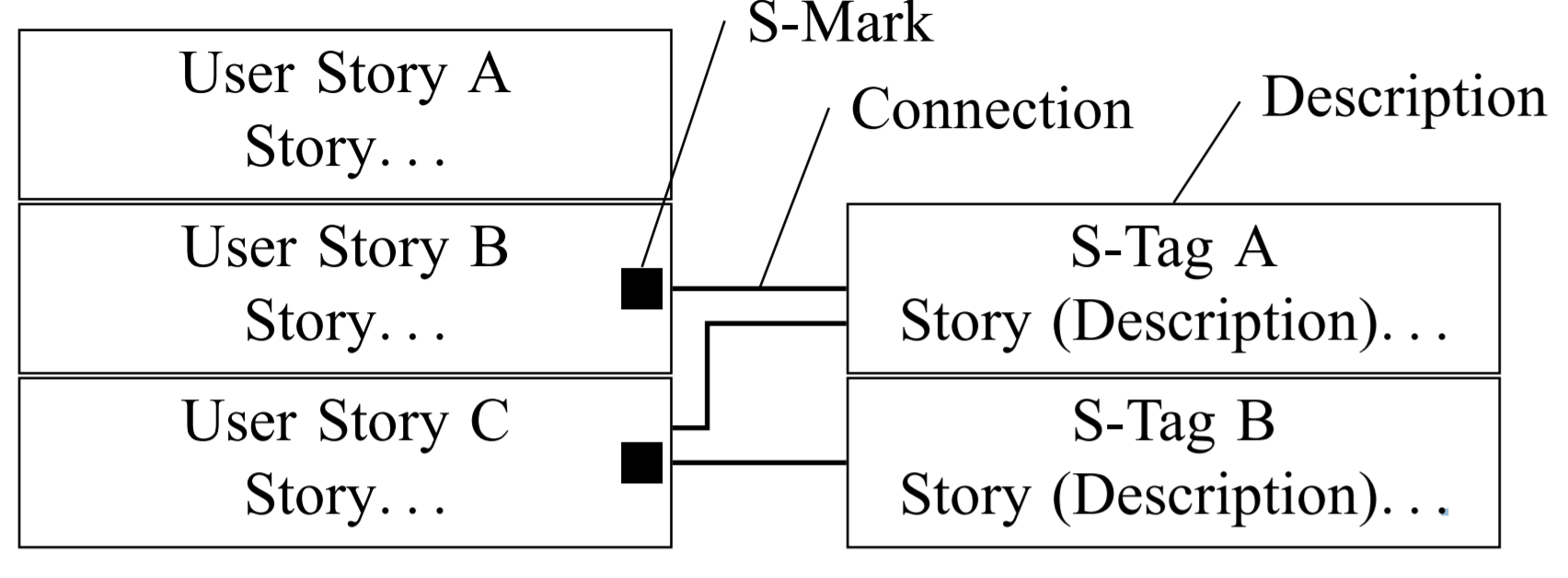}
	\caption[S-Tags: Traceable Security Concerns]{\label{fig:s-tags}} S-Tags. S-Tags function as tasks representing a security concern of a user story. S-Tags are connected to their related user stories through S-Marks, allowing a User Stories  security concerns to be tracked. User stories can have multiple S-Tags, and a single S-Tag may be related to several user stories. Once created, S-Tags are put into the backlog and completed normally. If a user story is added to a sprint, its related S-Tags must be added as well. \cite{RN52}.
\end{figure}

Munich IT Security Research Group revised S-Scrum by adding a connection between security tasks and related user stories \cite{RN52}. They called security tasks "S-Tags" (see: Figure \ref{fig:s-tags}). S-Tags are extracted from user stories like Security User Stories, but remain connected to the original story through "S-Marks", allowing developers to track security-related concerns for each user story. User stories can have multiple S-Tags, and vice versa \cite{RN52}. S-Tags are placed into the backlog and moved through the development process like ordinary stories. However, if a user story is added to a sprint, its related S-Tags must be added as well, keeping security a constant priority \cite{RN52}. S-Tags improve S-Scrum by improving the visualization and traceability of security within the context of Scrum's user stories. The Munich IT Security Research Group also performed an evaluation of their revised S-Scrum with a team of sixteen developers, and found that security of developed software was improved \cite{RN52}. 

Using the lessons from these versions of a security-focused Scrum, this paper attempts to translate them to Kanban and create a generalized framework for software focuses. 

\section{Kanban + X}

\subsection{Translation: Kanban + Security}

\begin{figure*}
	\centering
	\includegraphics[width=40pc]{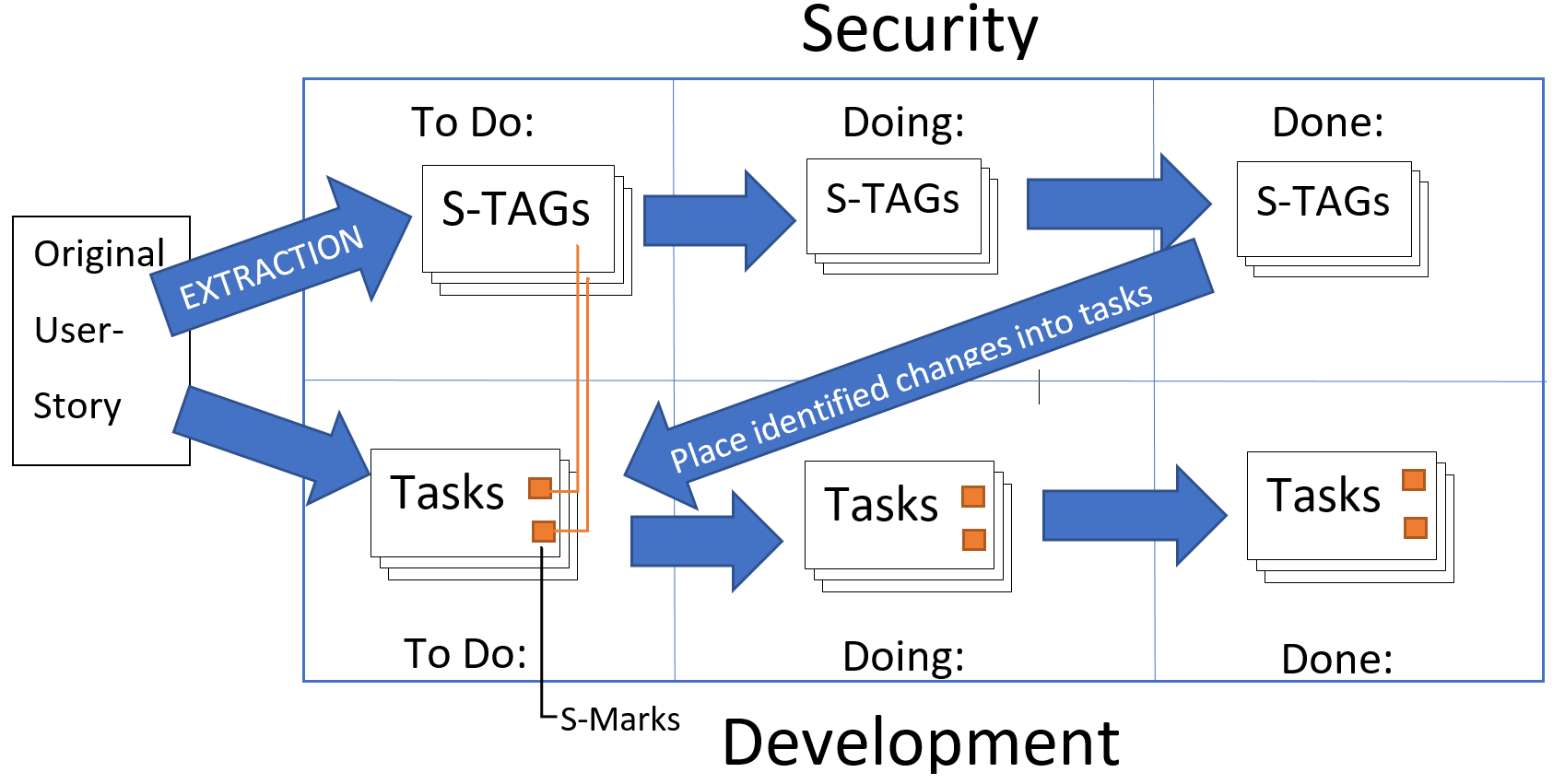}
	\caption[Kanban+Security]{\label{fig:kanbans}} The Kanban+Security work-flow: a translation of S-Scrum concepts into Kanban. First, a second board, Security, is added as a parallel iterative cycle like in ScrumS \cite{RN14}. Security related tasks, called S-Tags \cite{RN52} are extracted from user stories and placed on the Security board. Tasks are connected to one or more S-Tags through S-Marks for traceability \cite{RN52}. When tasks are started, their related S-Tags are started as well \cite{RN52}, contributing to the limit on in-progress tasks. As S-Tags are completed, security concerns are analyzed and needed changes are put into the beginning of the development board \cite{RN68}.
\end{figure*} 

To create a generalized framework for focused improvements, we will first translate the same security improvements to Kanban. Following the same steps as the S-Scrum researchers, we need to:

\begin{enumerate}
	\item Incorporate steps for security analysis and design into the Kanban work flow \cite{RN68}
	
	\item Place these steps into a parallel iterative cycle to be completed alongside development \cite{RN14}
	
	\item Visualize and track security concerns in "S-Tags" tied back to the original user stories. \cite{RN52}
\end{enumerate}

The results of these Kanban modifications can be seen in Figure: \ref{fig:kanbans}. We call this security-leveraged version Kanban + Security. First, we added an additional kanban board, called "Security" to act as the parallel iterative cycle found in ScrumS \cite{RN14}. This board acts like an ordinary Kanban board. Security tasks are still moved \textit{as needed} to reduce waste, and contribute to the limit on tasks "in progress" to keep down inefficiency from switching tasks \cite{RN18}. As in ScrumS, security tasks are extracted from user stories and put into the parallel iterative loop \cite{RN14}. As security tasks are moved through "doing", the analysis and design steps described in S-Scrum are completed \cite{RN68}. Needed changes are fed into the beginning of the development board, like in ScrumS \cite{RN14}, and are implemented normally. The security board's tasks are S-Tags, which are unchanged from their Scrum versions \cite{RN52}, increasing the visibility and traceability of security concerns. 

In this way, we have translated the security-focus of S-Scrum and its modifications to Kanban. Since it uses the same concepts, we expect similar benefits to the security of projects.

\subsection{Generalization}

\begin{figure*}
	\centering
	\includegraphics[width=40pc]{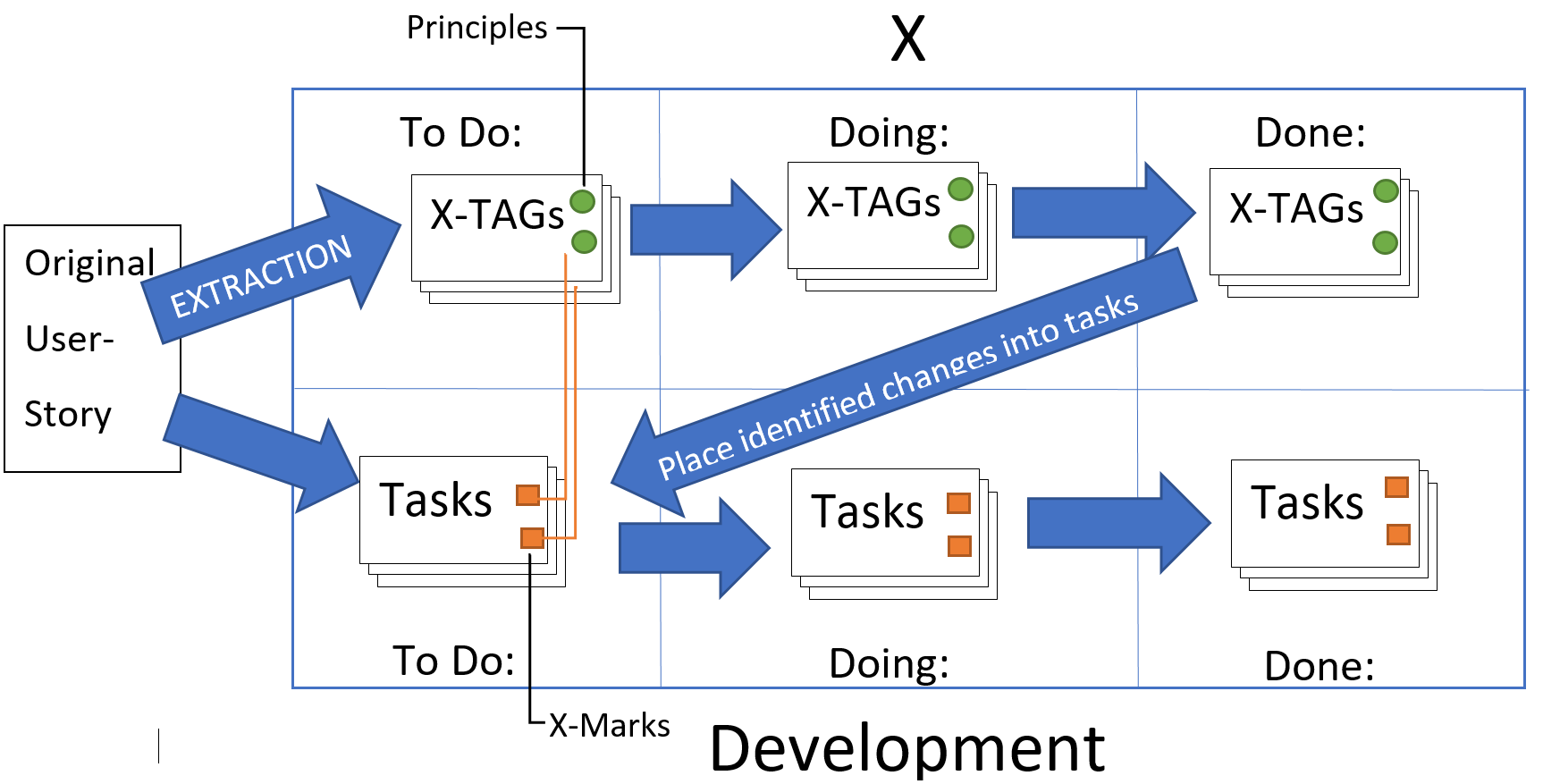}
	\caption[Kanban+X: A Generalized Framework]{\label{fig:kanbanx}} The Kanban + X framework work-flow. As a generalized version of Kanban+Security (See: Figure \ref{fig:kanbans}), the structure, appearance, and workflow are almost identical. Rather than Security or S-Tags, the second "X" board and "X-Tags" track whatever focus the project is being leveraged towards. X-Tags are also connected to one more principles, which are set by the development team and define the focus. 
\end{figure*}

Incorporating a specific focus does not only apply to security, but can be generalized to allow \textit{any} focus on software projects. This generalization is simple, and the results can be seen in Figure \ref{fig:kanbanx}. We call this focus-leveraged modification to Kanban Kanban + X, with "X" being whatever focus the project is being leveraged towards.

Kanban + X works mostly the same way as Kanban + Security. Focus-related tasks are extracted from the original user story and tracked on an additional Kanban board, functioning as a parallel iterative loop. Here, focus-related tasks are analyzed; Changes are designed based on said analysis and fed back into the original board. Most differences between Kanban + Security and Kanban + X are minor. References to security, such as "Security" board, "S-Tags", and "S-Marks" are replaced with "X" board, "X-Tags", and "X-Marks", with "X" standing-in for whatever focus Kanban is being leveraged towards.

The largest change for Kanban + X is the addition of "principles" (See: Figure \ref{fig:kanbanx}). Here "principles" refers to the set of good practices, values, and quality measures which make up a specific focus, as agreed upon by the team. For example, in security, the "principles" would be the concepts of risk and vulnerabilities \cite{RN14}. Principles increase the applicability of Kanban + X because any focus can be created by identifying the principles that define it. The principles are displayed and continuously reviewed and revised by the team to improve the focus-leveraging over time. During development, principles are referenced during creation of X-Tags, and the analysis and design of focus-required changes, thereby keeping X board progress within the agreed-upon definition of the focus. Every X-Tag is connected to one or more principles, in much the way Tasks are connected to X-Tags with X-Marks (See: Figure \ref{fig:kanbanx}). This way, each X-Tag has a visible indication of \textit{how} that task supports the focus.

\begin{figure*}
	\centering
	\includegraphics[width=40pc]{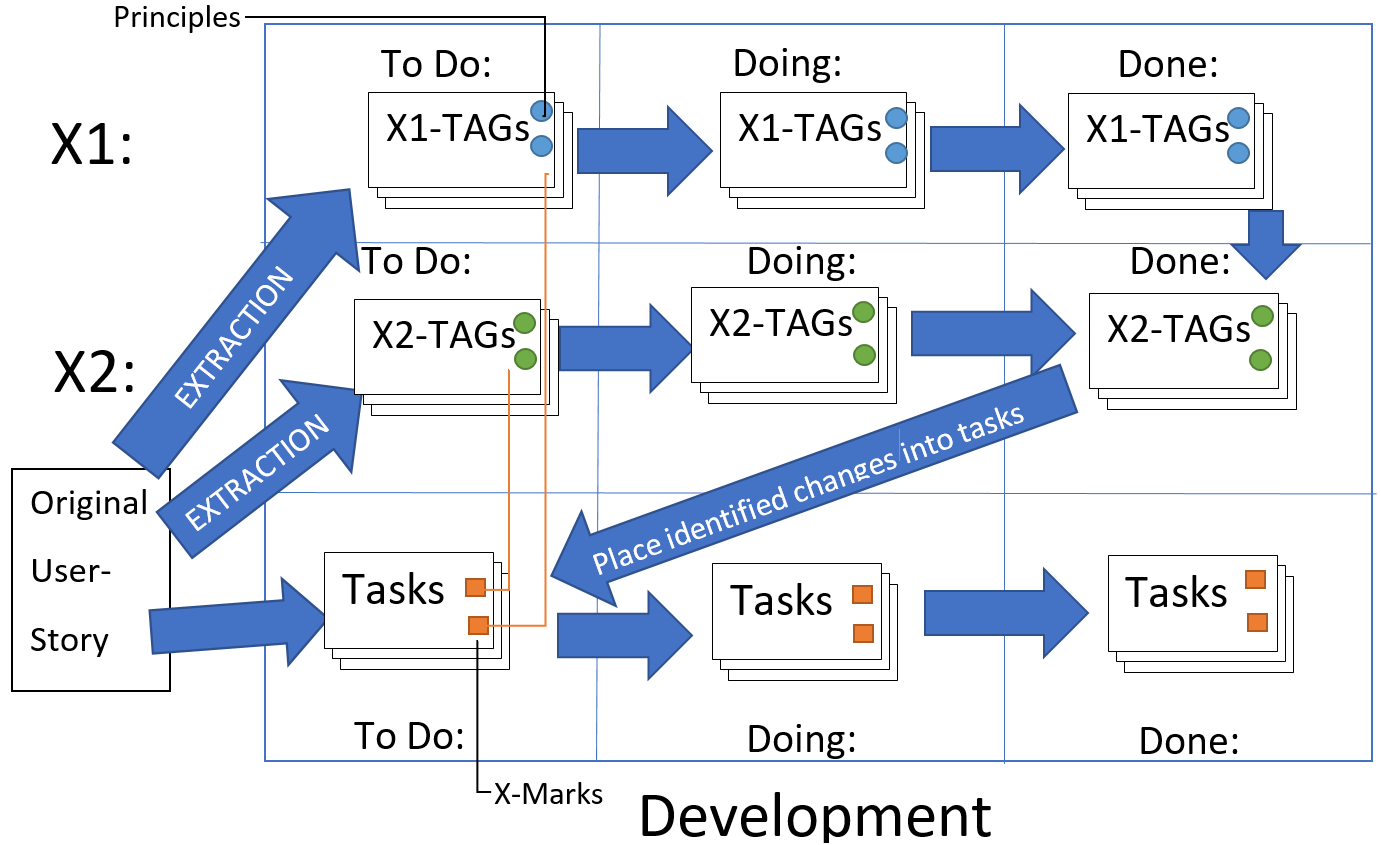}
	\caption[Kanban+X With Multiple Focuses]{\label{fig:multiplex}} Using Kanban+X with multiple focuses. Each focus has an X board which serves as a parallel iterative loop (See Figure: \ref{fig:kanbanx}). The X-Marks attached to Tasks may connect to any X-Tag from any board. X-Tags for each focus are extracted from the original user story according to principles for each focus, as set by the team. Each of X-Tag is analyzed, and changes are designed according to each focus's principles. Needed changes from each focus are fed into the development board together.
\end{figure*}

Furthermore, Kanban + X allows for the leveraging of multiple focuses at once (See: Figure \ref{fig:multiplex}). This is done by stacking one X board on top of another, each tracking progress towards a particular focus. Since X boards are parallel iterative loops, they function independently. X-Tags for each focus are extracted from user stories using each focus's principles. X-Tags are individually analyzed on their respective boards, and necessary changes from each focus are fed into the development board together. Obviously, leveraging for multiple focuses will also require additional balancing from the development team, but Kanban + X at least supports incorporation of multiple focuses for complex project. 

\subsection{Examples:}

To demonstrate the use of Kanban + X, we will apply the framework for leveraging towards Sustainability and Performance.

\subsubsection{Kanban + Sustainability}

We define sustainability as the ability of a software project to be developed and supported for its intended lifespan \cite{RN47}. This focus is important for large or complex project, which commonly fail due to non-sustainable practices \cite{RN47}, buildup of technical debt \cite{RN51}, and loss of knowledgeable team-members \cite{RN53}.

To apply Kanban + X to sustainability, we first replace generalities. The X board, X-Tags, and X-Marks are all renamed for sustainability and used for the parallel iterative loop shown in Figure \ref{fig:kanbanx}. Sustainability-Tags are extracted from user stories and completed based on the agreed-upon principles for sustainable development. We suggest three such principles based on published research on sustainable development:

\begin{enumerate}
	\item Team Code Ownership. 
	
	Code should be understood and contributable by all team members. This reduces risk of knowledge loss when develops leave, which can be devastating to sustainability \cite{RN53}. Furthermore, strong feelings of code ownership has been shown to increase quality and modifiability of the code \cite{RN49}. Sustainability-Tags related to Team Code Ownership may include knowledge sharing \cite{RN53}, requiring members to experience unfamiliar areas of code \cite{RN48}, verifying that code meets members' quality standards \cite{RN53}, and reviewing the team  for cohesion \cite{RN53}.
	
	\item Manage Technical Debt. 
	
	Technical debt is a trade-off: a short-term solution in exchange for needed refinement later. While useful, excessive debt can overwhelm a project \cite{RN51}. This principle refers to making good debt decisions, managing accrued debt, and eliminating unnecessary or useless debt. This principle's goal is to reduce technical debt, and make the remaining debt more useful. Related Sustainability-Tags may include documenting and tracking accrued debt, evaluating the origin and value of a debt decision, estimating debt repay-ability, and removing excess debt \cite{RN51}.
	
	\item Preventative Maintenance
	
	This principle refers to regularly pausing feature-production to catch up on overlooked or maintenance work \cite{RN47}. "Maintenance" includes many software engineering practices, such as automated tests, documentation, and refactoring, which improve long-term productivity but are often overlooked in favor of features \cite{RN47}. Related Sustainability-Tags may include writing and updating automatic tests,  writing \textit{useful} documentation \cite{RN50}, removing dependencies \cite{RN47}, and fixing low-quality code.
	
\end{enumerate}

By integrating these principles into the creation and analysis of Sustainability-Tags, changes for improved sustainability can be fed into the development cycle of the project, increasing sustainability.

\subsubsection{Kanban + Performance}

By "Performance", we refer to High Performance Computing, which is software's ability to complete complex programs efficiently, reliably, and quickly using multiprocessing. Performance is especially important for scientific research involving complex, highly computational simulations \cite{RN60}. Performance is challenging because it comes at the expense of other quality measures, such as readability, portability, etc \cite{RN61}. Because of the trade off, and research typically prioritizing results over software quality, most software engineering tools and practices are not used in HPC software development \cite{RN60}. Therefore, a development methodology leveraged towards Performance is desirable within the field \cite{RN58}.

We apply Kanban + X to performance much like Kanban + Sustainability. The "X" generalities are replaced with references to performance. Performance-Tags are extracted from user stories and completed in the parallel iterative loop of the Performance Board (see: Figure \ref{fig:kanbanx}). Performance's unique challenges are handled through agreed-upon principles, used for extracting and completing performance-tags. We have created three such principles based on the scientific goals of HPC and the unique trade-off of performance:

\begin{enumerate}
	\item Resource Management
	
	This principle refers to the primary action of HPC: the reduction of run time through efficient use of a computer's limited resources. This includes memory allocation, hardware optimization, and multiprocessing \cite{RN67}. Through this principle, tasks are analyzed for their efficiency, and changes to directly improve code performance are found. Related Performance-Tags may include documenting performance results, analyzing code for inefficiency, and optimizing code for specific hardware.
	
	\item Purpose
	
	This principle handles the trade off between performance and other quality measures. Its goal, in the style of Kanban, is to avoid unnecessary work by only optimizing code \textit{as needed}. Code should only be changed for performance if the resulting performance increase is worth the cost to readability, portability, etc, and if increasing a piece of code's performance would be significant to the overall performance of the software. Related Performance-Tags may include analyzing the cost of performance increases, balancing performance with other concerns, and determining whether a piece of code is worth optimizing. 
	
	\item Verifiability
	
	Due to the size and complexity of their calculations, HPC software is often difficult to verify. Scientific simulations especially may be expensive, dangerous, or impossible to physically replicate \cite{RN61}. Paradoxically, verifying results is extremely important to scientific research \cite{RN60}. This principle seeks to address this concern and improve HPC verifiability. Related Performance-Tags include analyzing verifiability of certain tasks, and improving verifiability where possible.
	
\end{enumerate}

Through these principles, Performance-Tags are created and completed, feeding changes for improved performance into the development cycle of the project. Kanban + Performance offers an agile development methodology for HPC while maintaining the emphasis on performance, which could be very useful for the community \cite{RN58}.

Specific principles of focus-leveraged development cycles will be different depending on the team and project. To be truly effective, the process should be further refined by practice and experience over time. Through the examples of Kanban + Sustainability and Kanban + Performance, we have demonstrated how the Kanban + X framework can be applied to \textit{any} focus.

\section{Conclusion}

As the security community has demonstrated, leveraging agile development processes towards a focus is beneficial to software projects for which the focus is valuable. Kanban + X provides a generalized framework for leveraging Kanban towards any particular focus, or combination thereof. We believe Kanban + X could be a useful tool for projects with a high-priority focus, such as sustainability or performance. By taking advantage of focus-leveraged development, such as Kanban + X, software projects can be more productive towards their specific goals.


\section*{Acknowledgment}

This material is based upon work supported by the National Science Foundation under Grant No. 1148188.

This material is based upon work funded by the U.S. Department of Energy Office of Science, Advanced Scientific Computing Research and Biological and Environmental Research programs. We thank program managers Thomas Ndousse-Fetter, Paul Bayer, and David Lesmes for their support.



\bibliographystyle{IEEEtran}
%
\bibliography{KanbanXBib}

\end{document}